\documentclass[sigconf]{acmart}

\usepackage{adjustbox}
\usepackage{algorithmic}
\usepackage{caption}
\usepackage{multicol}
\usepackage{import}
\usepackage{float}
\usepackage{algorithmic}
\usepackage{graphicx}
\usepackage{multirow}
\usepackage{tabularx}
\usepackage{textcomp}
\usepackage{xcolor}
\usepackage{xspace}
\usepackage{booktabs}
\usepackage{csquotes}
\definecolor{light-gray}{gray}{0.95}
\definecolor{pgreen}{RGB}{5,205,107}
\definecolor{pblue}{RGB}{2,154,223}
\usepackage{multirow}
\usepackage{listings}
\usepackage{algorithm}
\usepackage{optidef}
\definecolor{graphqlpink}{RGB}{236,29,151}
\definecolor{rubyred}{RGB}{169,20,1}
\definecolor{fakerred}{RGB}{222,63,36}
\definecolor{commentcolor}{HTML}{626262}
\definecolor{pteal}{RGB}{0,128,128}
\definecolor{stringcolor}{HTML}{17c6a3}
\definecolor{keywordcolor}{HTML}{dd2867}
\definecolor{classcolor}{HTML}{1290c3}
\definecolor{numbercolor}{HTML}{6897bb}
\definecolor{dim}{rgb}{0.55, 0.57, 0.67}
\usepackage{stackengine}
\usepackage{framed}

\AtBeginDocument{%
  \providecommand\BibTeX{{%
    \normalfont B\kern-0.5em{\scshape i\kern-0.25em b}\kern-0.8em\TeX}}}

\newcommand*\openquote{\makebox(12,-5){\scalebox{2}{``}}}
\newcommand*\closequote{\makebox(12,-5){\scalebox{2}{''}}}
\colorlet{shadecolor}{BurntOrange!15}

\makeatletter
\newif\if@right
\def\shadequote{\@righttrue\shadequote@i}
\def\shadequote@i{\begin{snugshade}\begin{quote}\openquote}
\def\endshadequote{%
  \if@right\hfill\fi\closequote\end{quote}\end{snugshade}}
\@namedef{shadequote*}{\@rightfalse\shadequote@i}
\@namedef{endshadequote*}{\endshadequote}
\makeatother

\usepackage{listings}
\lstdefinestyle{ruby}{
    language={ruby},
    basicstyle=\ttfamily\footnotesize, 
    morekeywords={},
    keywordstyle=\bfseries\color{rubyred},
    emph={Faker, Company, Internet, Color, Hacker, Hipster, Markdown, Internet, Books, Dune},
    emphstyle={\bfseries\itshape\color{fakerred}},
    stringstyle=\bfseries\color{teal},
    commentstyle=\itshape\color{fakerred}
}

\lstdefinestyle{js}{
    basicstyle=\ttfamily\footnotesize, 
    morekeywords={console, require},
    keywordstyle=\bfseries\color{rubyred},
    emph={},
    emphstyle={\bfseries\itshape\color{fakerred}},
    stringstyle=\bfseries\color{teal},
    commentstyle=\itshape\color{fakerred}
}

\newcommand{\lstbg}[3][0pt]{{\fboxsep#1\colorbox{#2}{\strut #3}}}
\definecolor{codegreen}{rgb}{0,0.6,0}
\lstdefinelanguage{diff}{
    basicstyle=\ttfamily\scriptsize,
	morecomment=[f][\color{red}]{---}, 
	morecomment=[f][\color{codegreen}]{+++},
	morecomment=[f][\lstbg{red!20}]{-\ },
	morecomment=[f][\lstbg{green!20}]{\ +\ },
	morecomment=[f][\color{blue}]{@@},}
	
\makeatletter
\lst@AddToHook{OnEmptyLine}{\addtocounter{lstnumber}{-1}}
\makeatother

\usepackage[framemethod=tikz]{mdframed}
\mdfdefinestyle{mpdframe}{
    frametitlebackgroundcolor   =BurntOrange!15,
    frametitlefont              =\itshape,
    frametitlerule              =true,
    roundcorner                 =1pt,
    middlelinewidth             =1pt,
    innermargin                 =0.1cm,
    outermargin                 =0.1cm,
    innerleftmargin             =0.1cm,
    innerrightmargin            =0.1cm,
    innertopmargin              =0.1cm,
    innerbottommargin           =0.1cm,
    linecolor                   =BurntOrange,
    skipabove                   =6pt
}

\newcommand{\revised}[1]{\textcolor{black}{#1}}

\newcommand{\faker}{\texttt{faker}\xspace}
\newcommand{\lolcommits}{\texttt{lolcommits}\xspace}
\newcommand{\volkswagen}{\texttt{volkswagen}\xspace}

\AtBeginDocument{%
  \providecommand\BibTeX{{%
    \normalfont B\kern-0.5em{\scshape i\kern-0.25em b}\kern-0.8em\TeX}}}

\copyrightyear{2024}
\acmYear{2024}
\setcopyright{rightsretained}
\acmConference[ICSE-SEIS'24]{Software Engineering in Society}{April 14--20, 2024}{Lisbon, Portugal}
\acmBooktitle{Software Engineering in Society (ICSE-SEIS'24), April 14--20, 2024, Lisbon, Portugal}\acmDOI{10.1145/3639475.3640099}
\acmISBN{979-8-4007-0499-4/24/04}

\begin{document}

\title[With Great Humor Comes Great Developer Engagement]{With Great Humor Comes\\ Great Developer Engagement}





\author{Deepika Tiwari, Tim Toady, Martin Monperrus, and Benoit Baudry}
\email{deepikat@kth.se, toady@eecs.kth.se, monperrus@kth.se, baudry@kth.se}
\affiliation{ 
\institution{KTH Royal Institute of Technology}
\city{Stockholm}
\country{Sweden} 
}
    
\renewcommand{\shortauthors}{Tiwari, Toady, Monperrus, and Baudry}

\begin{abstract}
The worldwide collaborative effort for the creation of software is technically and socially demanding.
The more engaged developers are, the more value they impart to the software they create. 
Engaged developers, such as Margaret Hamilton programming Apollo 11, can succeed in tackling the most difficult engineering tasks.
In this paper, we dive deep into an original vector of engagement -- humor -- and study how it fuels developer engagement.
First, we collect qualitative and quantitative data about the humorous elements present within three significant, real-world software projects: 
\faker, which helps developers introduce humor within their tests; 
\lolcommits, which captures a photograph after each contribution made by a developer; and 
\volkswagen, an exercise in satire, which accidentally led to the invention of an impactful software tool.
Second, through a developer survey, we receive unique insights from $125$ developers, who share their real-life experiences with humor in software.

Our analysis of the three case studies highlights the prevalence of humor in software, and unveils the worldwide community of developers who are enthusiastic about both software and humor.
We also learn about the caveats of humor in software through the valuable insights shared by our survey respondents. 
We report clear evidence that, when practiced responsibly, humor increases developer engagement and supports them in addressing hard engineering and cognitive tasks.
The most actionable highlight of our work is that software tests and documentation are the best locations in code to practice humor.
\end{abstract}





\maketitle

\section*{Lay Abstract}
Modern software applications are built piecemeal.
They are composed of a large pool of contributions made by individual software developers.
These developers spend long hours brainstorming with colleagues, solving challenging engineering problems, and writing code.
Software companies put different strategies in place to engage their developers and support them in coping with these hard tasks. 
In particular, some organizations and groups of developers decide to nurture the  creative and expressive aspects of software development.
In this work, we study a universal human trait that has been key for creativity since the dawn of civilization: humor. 
Humorous software code is a powerful deterrent to boredom.
It can uplift the mood of the developer who writes it, as well as a peer, separated by time and space, who chances upon this nugget in the codebase.
For example, Leia Organa is as perfect as John Doe for a placeholder customer name while testing an e-commerce application.
A gallery of photographs can serve as a cheerful journal of the Wednesdays spent hacking away with a colleague.
Software may also act as a humorous, satirical piece on socio-political issues.
In this paper, we discover the different ways in which software can be a source of developer amusement and engagement, and help foster a sense of community among them.

\begin{figure}[H]
\centering
\href{https://www.explainxkcd.com/wiki/index.php/974:_The_General_Problem}{\includegraphics[width=\columnwidth]{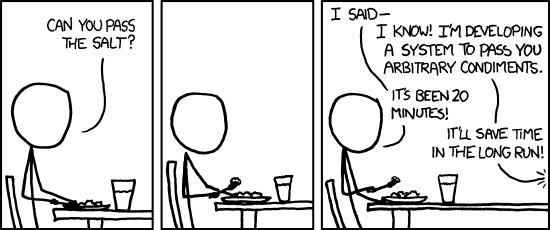}}
\caption{Comic strips are classics in software engineering~\cite{barros2017use}. Credits: \href{https://xkcd.com/}{xkcd}.}
\label{fig:xkcd}
\end{figure}
\section{Introduction}
The development of high quality software is certainly no laughing matter.
In addition to core technical problems, a major challenge of software engineering is its highly collaborative nature, which requires care and engagement.
To achieve this, different organizations create a culture of engagement through various corporate strategies.
For example, OpenBSD developers celebrate each major release through a new song \footnote{\url{https://www.openbsd.org/lyrics.html}}. 
Bluechip Silicon Valley company Stack Overflow choreographed the Dance Dance Authentication scheme \cite{dancedance} as company-wide software security training. 
Guidelines for organizing release parties exist for celebrating major milestones at Ubuntu \footnote{\url{https://wiki.ubuntu.com/BuildingCommunity/RunningReleaseParty/}}, Fedora \footnote{\url{https://docs.fedoraproject.org/en-US/mindshare-committee/events/release-parties/}}, and Debian \footnote{\url{https://wiki.debian.org/ReleaseParty}}.
Across companies, developer communities thrive within dedicated online channels \cite{aniche2018modern}
and through the gamification of the software engineering process~\cite{stol2022gamification}.



A crucial cultural phenomenon that comfortably predates software by at least $35,000$ years \cite{polimeni2006first} is the practice of humor.
Its purpose is to enchant and engage its audience \cite{taylor2003subterranean, west2019reverse}, and foster community-building.
Humor is fundamental to the human experience \cite{zargham2023funny}.
Our enterprising species has incorporated it in most of its creative undertakings, i.e., our spoken and written words, our art, and even traditional engineering \cite{ogunlana2006effect}.
Yet, to the best of our knowledge, the interplay of humor and software engineering remains largely unexplored in the literature.

In this paper, we unveil the phenomenon of developers engaging with software engineering through humor.
We start with case study research, and discuss three eminent examples of real-world humor in open-source software: the \faker library to generate non-boring test data, a tool called \lolcommits that captivates developers with selfies, and a parodic software tribute for making all tests pass in the CI, aptly named \volkswagen.
We highlight the humorous elements within each of these projects through quantitative and qualitative insights.
Next, we conduct a survey, inviting developers to share their personal practice of, and experiences with, humor in the software they create. 

Our results distinctly show the nuances of humor within software.
The \faker ecosystem is continually evolving through new humorous material contributed by enthusiasts.
\lolcommits has been a handy tool for fostering collaboration, celebrating achievement, and reflecting on the passage of time.
In addition to a joke, the software engineering community has also gained a useful tool as a side-effect of \volkswagen. 
Overall, the community of developers who engage with the humorous elements within these projects is remarkable.
Furthermore, $100$ of the $125$ respondents of our survey testified to experiencing humor within software.
They share their valuable insights on practising humor responsibly \cite{peifer2012can}, and we learn that humor is most likely to be found within code comments and test inputs.

The work most closely related to our original inquiry is about team communication   
\cite{mendez2018open, hoffmann2022human}. 
There are rare studies on software-related humor, such as on \emph{Easter eggs} \cite{baudry2022long, lakier2022more}, and \emph{esolangs} \cite{temkin2017language}.
Humor and creativity are closely associated, and this also holds for software engineering \cite{GroeneveldLVA21}.
The novel aspect of this work is its focus on humor in programming and code, reconciling the  rigorous engineering endeavour with our human inclination towards humor.

We summarize our contributions as follows.
\begin{itemize}
    \item An exploration of the diverse facets of real-world humor in software projects.
    \item Actionable knowledge from developers on how to practice humor in their art, in order to create engaging developer communities.
    \item A call for the software engineering research community to take humor seriously.
\end{itemize}

In \autoref{sec:engagement-with-humor}, we dive deep into three illustrious examples of humor in software engineering.
Next, \autoref{sec:developer-survey} presents the first ever survey targeted towards understanding humoristic developer practices.
We discuss prior art on humor-related software in \autoref{sec:related-work}, before concluding this piece in \autoref{sec:conclusion}.








\section{Case Studies of Software Developer Engagement with Humor}\label{sec:engagement-with-humor}

\begin{table*}
\renewcommand*{\arraystretch}{1.6}
\centering
\caption{Humorous components within software projects can create developer engagement.
We illustrate this phenomenon through three popular, open-source projects on GitHub: \faker, \lolcommits, and \volkswagen.
For each project, we highlight the number of contributors (\textsc{\#Contributors}) and stargazers (\textsc{\#Stars}), which capture quantitative evidence of developers' engagement with these projects on GitHub. 
In the \textsc{Key Fact} row, we report  project-specific metrics reflecting its admiration among developers.
The \textsc{Quotes} have been sourced from developer forums, social networks, and personal websites.}\label{tab:funny-projects}
\resizebox{\textwidth}{!}{
\begin{tabular}{|l|r|r|r|}
\hline
 & \href{https://github.com/faker-ruby/faker}{\textbf{\faker}} & \href{https://github.com/lolcommits/lolcommits}{\textbf{\lolcommits}} & \href{https://github.com/auchenberg/volkswagen}{\textbf{\volkswagen}} \\ \hline
\textsc{\#Contributors} & $845$ & $57$ & $13$ \\ \hline
\textsc{\#Stars} & $10,882$ & $4,606$ & $12,716$ \\ \hline
\textsc{Key Fact} &
\begin{tabular}[c]{@{}r@{}} $141$ funny data generators \\ contributed by $200$ developers\end{tabular} 
& \begin{tabular}[c]{@{}r@{}}{Gem downloaded $266,365$ times,} \\ {Multiple video montages and} \\ {galleries, e.g., on YouTube}\end{tabular} & \begin{tabular}[c]{@{}r@{}}{Featured $23$ times}\\ {as top trending on GitHub}\\ {between 2015 and 2021}\end{tabular} \\ \hline
\textsc{Quotes} & 
\begin{tabular}[c]{@{}r@{}} \emph{``I love the faker gem. It's so} \\ \emph{useful when you want to} \\ \emph{populate a dev db."} \\ --- \href{https://www.reddit.com/r/ruby/comments/10wh1c4/comment/j7p446c/}{xutopia on Reddit}\\
\arrayrulecolor{orange}\hline
\emph{``Love using Faker to populate} \\ \emph{demos with ``real" content vs.} \\ \emph{just using [Lorem] Ipsum"} \\ --- \href{https://www.reddit.com/r/ruby/comments/xc6cd3/comment/io52rdd/}{Ecstatic-Leader485 on Reddit}\\
\hline
\emph{``[Faker] is a work of art"} \\ --- \href{https://twitter.com/lylo/status/1686494559721705472?s=20}{@lylo on Twitter / X} \\
\hline
\emph{``Part of your daily Rails}\\ \emph{testing diet"} \\ --- \href{https://www.reddit.com/r/ruby/comments/xc6cd3/comment/io427x2/} {Longjumping\_You\_1786 on Reddit}
\end{tabular} &
\begin{tabular}[c]{@{}r@{}} \emph{``I will enable this at home for my} \\ \emph{side-projects! This is really LOL.} \\
--- \href{https://news.ycombinator.com/item?id=6197933}{daGrevis on Hacker News} \\ 
\arrayrulecolor{orange}\hline
\emph{``I've been running lolcommits} \\ \emph{for 10 years [...]"} \\ --- \href{https://persistent.info/commits/}{Mihai, blog post} \\
\hline
\emph{``I turned on lolcommits so I'd have}\\ \emph{a record of my working and kept} \\ \emph{hacking away. lolcommits is} \\ \emph{seriously underrated [...]"} \\ --- \href{https://evantravers.com/articles/2019/11/21/making-a-book-using-ruby/}{Evan, blog post} \\
\hline
\emph{``This is a great idea - I wonder}\\ \emph{[why] I never thought of it"} \\ --- \href{https://news.ycombinator.com/item?id=6197933}{xmpir on Hacker News}
\end{tabular}
&
\begin{tabular}[c]{@{}r@{}} \emph{``I appreciate the author of this library}\\ \emph{to no end. There  is something to be}\\
\emph{said about software being made} \\ \emph{in jest."} \\ --- \href{https://news.ycombinator.com/item?id=18067861}{birdiesanders on Hacker News} \\
\arrayrulecolor{orange}\hline
\emph{``My god, this is hilarious"} \\ --- \href{https://news.ycombinator.com/item?id=18067861}{clircle on Hacker News}\\
\hline
\emph{``This is a work of genius. Best laugh}\\ \emph{I've had all week."} \\ --- \href{https://www.reddit.com/r/javascript/comments/3nvzbj/comment/cvrxk0u}{diatu on Reddit} \\
\hline
\emph{``This package is still one of my} \\ \emph{favorite things on the internet,} \\ \emph{ever."} \\ --- \href{https://news.ycombinator.com/item?id=28180352}{sensitive-ears on Hacker News}
\end{tabular}
\\ 
\hline
\end{tabular}
}
\end{table*}

On a typical workday, a developer  writes code to implement a new feature, or to fix a bug.
Additionally, a diligent developer writes tests to verify that the new code behaves as expected.
The next step is to \emph{commit} these changes to the project.
Finally, our developer pushes this commit to the remote server where the project resides.
After peer-review, all project-wide incoming commits trigger a sequence of events within the \emph{Continuous Integration} (CI) pipeline.
The most important event within this pipeline is to run the whole suite of tests in the project.
Only if the test suite executes successfully are the new changes merged into the project.
The process of coding, committing, testing, and integrating is perpetual.
In this section, we investigate how humor has made its way into every step of this cycle. 

We conduct case-study based research \cite{flyvbjerg2006five} and analyze three notable, open-source development tools, which introduce humor in tests (\autoref{sec:faking-libraries}), commits (\autoref{sec:lolcommits}), and CI (\autoref{sec:volkswagen}), respectively.
For each of these projects, we report quantitative data about its codebase and community engagement, as well as qualitative insights from our manual analysis of the project.
We also gather first-hand accounts of the significance of these humorous tools from interviews with developers who maintain or use them.

\subsection{Testing: Faking It While Making It}\label{sec:faking-libraries}

Academics as well as industry practitioners unanimously agree that testing is crucial to developing high quality software \cite{incrementtesting}.
Developers write tests to ensure that the features they have implemented within their system behave as expected, and that the introduction of new code does not introduce regressions.
Each test contains components called test inputs that bring the system to a testable state.
This involves initializing variables or objects, and setting up the resources required for the test, such as a database.
Test inputs are required to trigger, and consequently verify, different software behaviors, yet constructing good test inputs is known to be challenging and time-consuming \cite{choudhary2015automated}.
A solution to this problem is the use of data generators that synthesize realistic test inputs, such as a collection of customer contact details. These data generators are packaged in so-called  \emph{faking libraries}.
\faker is the gold standard for faking in Ruby \footnote{\url{https://rubygems.org/gems/faker}}.
It contains more than $20,000$ lines of Ruby code, and as reported in \autoref{tab:funny-projects}, has been starred by more than $10,000$ users on GitHub.
As of this writing, \faker provides $232$ object generators.
Each of these generators has meticulously been documented with examples.

\faker provides a plethora of fundamental data generators, such as \texttt{address}, \texttt{barcode}, \texttt{date}, and \texttt{lorem}.
In addition to  conventional generators, the $845$ contributors of \faker have deemed it necessary to include generators that support developers with humorous data, such as a \texttt{lebowski} quote string generator, or a \texttt{funny\_name} generator, to obtain \texttt{Ben Thair} or \texttt{Don Thatt} rather than  the boring suspects, \texttt{Jane Doe} or \texttt{John Doe}.
These extraordinary generators in \faker refer to cultural elements such as movies, TV shows, music, video games, sports, and books, among others. 
For example, \texttt{Faker::TvShows::Seinfeld.business} yields one of the $23$ Seinfeld-related business establishments, such as \href{https://youtu.be/Ugx06TlVyw4?si=kjOOlpz2fh_p3DHH}{Vandelay Industries} \footnote{\url{https://github.com/faker-ruby/faker/blob/main/lib/locales/en/seinfeld.yml}}.
\faker is meant to bring joy to test inputs.

In order to understand this humorous aspect of \faker, as well as the technical and creative challenges of maintaining a project of this stature, we interviewed Stefanni Brasil, a core maintainer of the project.
Stefanni attributes the vibrancy and allure of \faker to Matz, the creator of Ruby.
\emph{``Matz wants programmers to be \href{https://github.com/faker-ruby/faker/blob/13140408843a1713385829007c1cbfc7396628a5/lib/locales/en/quote.yml\#L79}{happy}."} \footnote{\url{https://youtu.be/oEkJvvGEtB4?t=1795}}
Speaking about generator contributions to \faker, Stefanni remarked, \emph{``I frequently review pull requests from people who want to add their favorite TV show,"} suggesting the emotional driver behind developer contributions. 
Indeed, through manual analysis, we determine that at least $141$ of the $232$ generators in \faker may be considered unconventional and humorous.
Over the $16$ years that \faker has been in active development, these generators have generously been contributed by $200$ open-source enthusiasts.
However, performance and engineering constraints must be considered before adding a new generator \footnote{\url{https://github.com/faker-ruby/faker/issues/2689}}.
For this reason, every contribution to \faker follows a template, is accompanied by relevant tests, documentation, and locale support, and undergoes a stringent review process.
Good humorous fakes  require real engineering. 

The continuous growth of the \faker codebase has also been a subject of discussion among developers, who have previously been conflicted about moving a basic family of generators into a \texttt{faker-lite} gem \footnote{\url{https://github.com/faker-ruby/faker/issues/1539}}.
They currently lack the data on how generators are used in the wild, which is necessary to ground the selection of a subset of generators for \texttt{faker-lite}.
Stefanni notes, 
\emph{``We only see the number of downloads, but we would really like to know how developers use the generators. Many of these uses may be in private projects, so there is no easy way to know."}
Besides, as highlighted by a developer in the GitHub conversation, providing the full suite of generators by default facilitates an easier introduction of fun into software tests.
In fact, such a split could complicate things for users.
Stefanni guarantees that the humorous generators within \faker are safe for the foreseeable future.

We query GitHub to investigate the real-world usage of \faker in open-source projects.
One notable client of the \faker library is a project called \texttt{Forem} \footnote{\url{https://github.com/forem/forem}}, which powers a social networking platform for developers.
It uses \faker to populate its application and database for test execution, such as with the file 
\href{https://github.com/forem/forem/blob/7e83f18eca805a052a34e7d7c41a264119765a55/db/seeds.rb}{\texttt{seeds.rb}}.
This file calls $10$ distinct generators of \faker.
We show an excerpt of this file in \autoref{lst:faker-in-forem}, where the database is being seeded with some fake \texttt{Comment}s and \texttt{Page}s.
Several attributes of the two entities, such as the \texttt{body\_markdown} of each \texttt{Comment} (line $5$), are produced using \faker generators.
The comments on lines $4$, $15$, and $18$ are sample outputs from the three showcased generators.
Note that the \texttt{title} of each \texttt{Page} (line $16$) could easily be a \texttt{Faker::Lorem.sentence}.
Instead, it is something smart a \texttt{Hacker} would say \footnote{\url{https://github.com/faker-ruby/faker/blob/main/lib/locales/en/hacker.yml}}.
Working on software projects for her day job in a consultancy firm, Stefanni shared how she uses \faker for testing.
\emph{``The good thing about testing is that we can be creative. I often use \texttt{cat} names. But my favorite is the \texttt{parks\_and\_recreation} generator. It is my favorite show."}

\begin{lstlisting}[style=ruby, showlines=true, label={lst:faker-in-forem}, caption={The developers of the \texttt{Forem} project seed their development database with some humor using \faker}., float]
 seeder.create_if_none(Comment, num_comments) do
   num_comments.times do
     attributes = {
       # hashtag hella art party.
       body_markdown: Faker::Hipster.paragraph( sentence_count: 1),
      ...
    }
    Comment.create!(attributes)
   end
 end
 ...
 seeder.create_if_none(Page) do
   5.times do
     Page.create!(
       # We need to override the haptic JBOD pixel!
       title: Faker::Hacker.say_something_smart,
       ...
       %*\href{https://youtu.be/j1epEtB0lVo}{\textcolor{rubyred}{\# The wise animal blends into its surroundings.}}*)
       description: Faker::Books::Dune.quote,
       ...
     )
   end
 end
\end{lstlisting}

\faker is only one member of a large and rich ecosystem of faking libraries.
The PERL \texttt{Data::Faker} module \footnote{\url{https://metacpan.org/pod/Data::Faker}} was the seed for this ecosystem, which ships with only $6$ conventional generators.
\faker, inspired by \texttt{Data::Faker}, brought the idea of humorous generators, in addition to the conventional ones ported from PERL.
This gem eventually triggered the development of other faking libraries, both in Ruby, such as \texttt{ffaker} \footnote{\url{https://github.com/ffaker/ffaker}}, as well as in other programming languages, including Java, Python, PHP, JavaScript, Haskell, and even C++.
Remixing the design of \faker, many of these libraries offer diverse sets of vanilla as well as unconventional generators.
For example, \texttt{java-faker} \footnote{\url{https://github.com/DiUS/java-faker}} has also introduced a new generator to produce quotes from the French comedy show \texttt{Kaamelott}.
We argue that the academic community can gain from paying more attention to faking libraries with respect to both testing and humor.

\begin{mdframed}[nobreak=true,style=mpdframe, frametitle=Summary of the \faker case study]
Hundreds of developers have made humorous contributions to faking libraries such as \faker.
Humor helps foster open-source collaboration beyond company walls.
With humorous faking, software testers worldwide are treated with a nugget while debugging a failing test.
\end{mdframed}

\subsection{Commits: LOLs All the Way Down}\label{sec:lolcommits}
A \emph{commit} can be considered as the atomic unit of contribution to a software project managed by a version control system.
A commit includes code changes, as well as a brief textual description of the changes.
Each commit may be likened to the snapshot of the project at a point in time, and is uniquely identifiable through its hash.
As an example, we can consider the \texttt{Forem} project discussed in \autoref{sec:faking-libraries}. With commit \href{https://github.com/forem/forem/commit/944991a432575b8386c22715722e6ddc28d48613}{\texttt{944991a}}, a \texttt{Forem} developer has implemented a new feature, updated related tests, and described these changes with a commit message.

The number of commits made by developers globally on a workday is likely to be in the millions \cite{zhao2017impact}.
Given the significance of commits, there is an ecosystem of utilities that aid developers in committing.
For example, commit messages may be linted to ensure that they abide by conventions with \texttt{commitlint} \footnote{\url{https://github.com/conventional-changelog/commitlint}}. The \texttt{git- secrets} \footnote{\url{https://github.com/awslabs/git-secrets}} tool scans commits to prevent credentials and secrets from accidentally making their way into the project codebase.
A new way to commit is to let generative AI handle the message, such as with \texttt{aicommits} \footnote{\url{https://github.com/Nutlope/aicommits}} or \texttt{opencommit} \footnote{\url{https://github.com/di-sukharev/opencommit}}.

Admittedly, commits drive software development.
However, they do not have to be mundane.
\texttt{gitmoji} \footnote{\url{https://github.com/carloscuesta/gitmoji}} serves as a handy tool for identifying the right emoji for a commit message, such as \texttt{:egg:} for a commit that adds or updates an Easter egg.
Another crowd favorite is \lolcommits, which has been downloaded upwards of $250,000$ times as of August $2023$ \footnote{\url{https://rubygems.org/gems/lolcommits/versions/0.16.3}}.
\lolcommits is a command-line utility that captures a portrait of a single developer or a group of developers, and attaches it to each commit made to a project.
\autoref{tab:funny-projects} presents some details on the \lolcommits project.
Nearly $60$ developers have contributed to it, while the number of stargazers of \lolcommits is more than $4,600$. 

As noted on its GitHub \texttt{README}, \lolcommits was originally a \emph{``joke project"} showcased at a Hack \&\& Tell meetup \footnote{\url{https://hackandtell.org/}}, by its creators Matthew R. and Matthew H.
It has since matured, and can now be installed on computers running all major operating systems.
When enabled for a project, \lolcommits adds a post-commit hook to its \texttt{.git} folder.
Then, each time a commit is made, \lolcommits triggers a capture from the camera installed on the device.
By default, each lolcommit includes the first $11$ characters of the commit SHA, as well as the commit message, typed out in meme-like typography.
Over the $12$ years since its inception, the contributors to \lolcommits have consistently provided new features, including different modes, typography, and additional plugins. 
\lolcommits is committed to developers; it also supports the generation of videos and animated GIFs, which are features much appreciated by its user base.

\begin{figure}
\centering
\href{https://youtu.be/dQw4w9WgXcQ}{\includegraphics[width=\columnwidth]{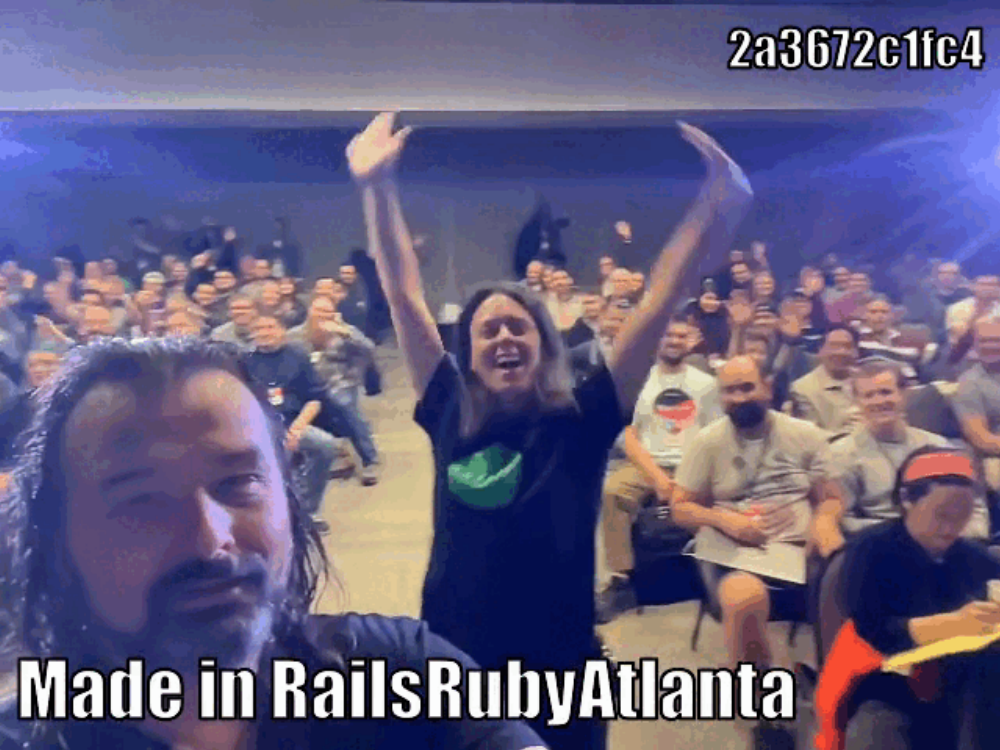}}
\caption{A crowd enjoying \lolcommits in April, 2023}
\label{fig:lolcommit}
\end{figure}

\lolcommits has a thriving community.
There are creative video montages featuring the \lolcommits of developers over the years \footnote{\url{https://youtu.be/mdzPI7Id840}}.
Several developers also mention using \lolcommits on their blogs and social pages.
We contacted three such ardent users of \lolcommits.
Selena Small and Michael Milewski are colleagues who pair-program together, give talks in conferences across the world, and participate in hackathons.
They often feature \lolcommits in their projects and mentorship material.
They also lolcommit live on stage, as part of their technical talks \footnote{\url{https://youtu.be/qCS1CYxOiEo?si=9kzesLgK3TNaCfi0}}.
Evan Travers is a developer who has used \lolcommits to document his progress on both professional and personal projects.
Speaking about how she started using \lolcommits, Selena recalled, \emph{``As pair programmers, Michael and I constantly share our ideas and setup with each other. That is how I learned about \lolcommits."}
Michael on the other hand happened to be walking by when a colleague at his previous job was making a lolcommit.
\emph{``I accidentally found myself in a lolcommit, and thought it was really fun. It works so well!"}
Evan discovered \lolcommits together with his colleague, the developer of a Ruby gem for Urban Dictionary \footnote{\url{https://rubygems.org/gems/urban/}}.
\emph{We were subscribed to all newsletters about Ruby gems. One of them mentioned \lolcommits.}

Attaching a photo to a commit may serve multiple purposes.
Evan recounts a challenging day at work trying to solve a Heisenbug \cite{musuvathi2008finding} which caused a broken build.
\emph{``It is nice to have a visual journal of how excited or frustrated you felt. While we were working on fixing that evasive bug, \lolcommits captured photos of colleagues looking over my shoulder, pointing at my screen. We reflected on them over pizzas."}
Michael and Selena believe \lolcommits can help keep things \emph{``fun and lighthearted"} while working seriously.
\emph{``You can pose for a photo with your colleagues, be creative, even use props. It is an immediate celebration of your work!"}
\autoref{fig:lolcommit} presents one of the \lolcommits authored by Selena and Michael, starring many excited participants.
Their own archive of \lolcommits features several colleagues, amounts to $6.5$ GB of disk space, and represents years of work fulfilment.
They add that incorporating live \lolcommits on stage during their talks \emph{``adds an element of fun to our extravagant, technical presentations."}
Selena and Michael also share their experience using \lolcommits while on a Rails Camp \footnote{\url{https://rails.camp/}}, a $52$-hour excursion in the woods to practice pair-programming.
Their goal was to have at least one significant commit with each of the $38$ participants during the camp.
Selena found that \lolcommits was instrumental in this context,
\emph{``We used it for onboarding new developers."}
Towards the end of the camp, most people were excited about taking post-commit photographs \footnote{\url{https://github.com/failure-driven/blog/blob/master/content/rails-camp/commit-31-given-when-then-steps.md\#lolcommit}}.
\emph{``Pair-programming is a good way to understand how people work, and \lolcommits was the cherry on the cake!"}

\begin{mdframed}[nobreak=true,style=mpdframe, frametitle=Summary of the \lolcommits case study]
Developers make millions of commits everyday.
Introducing humor through tools such as  \lolcommits is rewarding for developers, and contributes to a celebration of the work done.
In addition to journaling with personal selfies, it fosters relationships with colleagues by encouraging in-person, group \emph{usies}.
\end{mdframed}

\subsection{CI: But It Works on My CI}\label{sec:volkswagen}
Every so often, a large corporation receives some bad press.
Volkswagen made headlines in $2015$, when it was discovered that they were not completely honest about their manufactured automobiles.
The software embedded within their diesel vehicles reported lower emissions of environmental pollutants, specifically during test sessions in the laboratory \cite{schiermeier2015science}.
This discovery had significant consequences, and key decisions were made such that car manufacturers undergo more scrutiny.

The Volkswagen story resonated with developers who earnestly test their projects within Continuous Integration (CI) pipelines.
Some enterprising developers saw the Volkswagen headline as an opportunity to implement their own \emph{defeat devices}.
Such a device can tinker with test harnesses, tricking them into believing that all tests in the test suite pass, thus circumventing the CI to return the coveted green \texttt{:check\_mark:}.
The first software project to achieve this challenging technical feat was \texttt{phpunit-vw}, built for PHP projects \footnote{\url{https://github.com/hugues-m/phpunit-vw}}.
Since then, such defeat devices for the CI have been implemented across multiple programming languages, including JavaScript, Rust, Ruby, Java, and even C++ \footnote{\url{https://github.com/WyriHaximus/awesome-volkswagen}}.
To counter these, there also exists hoaxwagen, which detects if test results in the CI are being manipulated \footnote{\url{https://github.com/CleanCode-Group/hoaxwagen}}.

Developers Kenneth Auchenberg and Thomas Watson implemented \volkswagen in JavaScript.
During our interview with Thomas, he recalled, \emph{``We were inspired by the PHP version, and wanted to see if it could be done in JavaScript."}
They leverage a technique called monkey patching, which dynamically modifies a function at runtime \cite{pfretzschner2017identification}.
\volkswagen hooks into Node and patches the \texttt{require} function.
This modified \texttt{require} function links to a version of the test framework, which itself is modified to return a successful execution status for all tests. 
As a safeguard, \volkswagen also checks that this monkey patching is activated only in the CI, and that the behavior of the tests is not changed if they are run locally.
Thomas clarified, \emph{``On the surface it might seem easy, but it is a challenge. You have to be aware of all testing frameworks, and also make it work with the whole, messy JavaScript ecosystem."}
The first release of \volkswagen was \emph{``urgent,"} being rolled out within $24$ hours after development started.
Currently, \volkswagen supports at least $18$ different CI servers, and is capable of intercepting multiple JavaScript testing frameworks.

\begin{lstlisting}[style=js, showlines=true, label={lst:is-ci}, caption={The development of \volkswagen triggered the development of \texttt{is-ci}, which has since found legitimate use cases.}, float]
const isCI = require('is-ci');

// from %*\href{https://github.com/arduino/arduino-ide/blob/73ddbefc3e22c65d00ff59cade57fda14af01818/electron-app/scripts/notarize.js#L8}{arduino/arduino-ide}*)
if (!isCI) {
  console.log('Skipping notarization: not on CI');
  return;
}%*\par\noindent\dotfill*)

// from %*\href{https://github.com/salesforce/observable-membrane/blob/bd74b5c1d12d24a4cbddb484d5cae3639f22be2a/scripts/release/publish.js#L12}{salesforce/observable-membrane}*)
if (!isCI) {
  console.error('This script is only meant to run in CI.');
  process.exit(1);
};%*\par\noindent\dotfill*)

// from %*\href{https://github.com/videojs/video.js/blob/38b165d4d79528a73a97cee15776c908425ddb8b/rollup.config.js#L91}{videojs/video.js}*)
const progress = () => {
  if (isCI) {
    return {};
  }
  return progressPlugin();
}
\end{lstlisting}

Thomas admits that \volkswagen is a meant as joke, and does not envision a direct use case for it.
Yet, \volkswagen is a good candidate for a tool that attacks the supply chain by making tests pass, despite the introduction of malicious code \cite{torres2019toto}.
While working on \volkswagen, Thomas implemented a tool called \texttt{is-ci}, which identifies if the execution environment is in fact a CI server.
He released \texttt{is-ci} as an independent package on the \texttt{npm} registry \footnote{\url{https://www.npmjs.com/package/is-ci}}.
Contrary to \volkswagen, \texttt{is-ci} is profoundly useful.
We highlight three of the thousands of open-source usages of \texttt{is-ci} in \autoref{lst:is-ci}.
The \texttt{arduino-ide} project from \texttt{arduino} (lines $2$ to $6$), and the \texttt{observable-membrane} project from \texttt{salesforce} (lines $8$ to $12$), rely on the output of \texttt{is-ci} to perform certain actions only during execution on the CI.
Conversely, the \texttt{video.js} project (lines $14$ to $20$) uses \texttt{is-ci} to skip an operation if the execution environment is the CI.

\volkswagen has had two worldwide effects.
First, it became a project of Internet fame with its purely humorous nature, having been featured on the daily list of top-trending projects on GitHub $23$ times between October $2015$ and August $2021$ \footnote{\url{https://github.com/larsbijl/trending\_archive}}, as highlighted in \autoref{tab:funny-projects}.
\emph{``It blew up! Kenneth, whose GitHub profile hosts \volkswagen, became the top-ranked developer on GitHub while we were trending for the first time,"} Thomas recalled.
Second, it triggered the implementation and release of the open-source package \texttt{is-ci}, which responds to an essential CI/CD use case.

\begin{mdframed}[nobreak=true,style=mpdframe, frametitle=Summary of the \volkswagen case study]
All components of software development can be the subject of humor.
For example, thousands of developers have shared a joke about continuous integration.
Software developed purely for fun may even lead to the inception of a truly impactful tool, such as \texttt{is-ci}.
\end{mdframed}

\subsection{Takeaways from the Case Studies}
Our three case studies, \faker, \lolcommits, and \volkswagen, highlight the intricacies of humor within software.
Through \faker, we see that humor can be incorporated within testing to engage developers and testers.
The \lolcommits project gives developers the opportunity to celebrate their development journey, and their ongoing collaboration with colleagues in a playful manner.
\volkswagen is a détournement of a societal issue, unexpectedly leading to a popular and useful open-source package.
\revised{Each of these projects addresses a fundamental activity within software engineering, while offering an opportunity for developers to express their creativity and engage with their peers.
The billions of software projects that power our world are diverse, yet each may have numerous treasures planted by contributors, waiting to be unearthed by unsuspecting wayfarers.}
The community of developers who engage with humorous practices within software development is large, to an extent which is not known in the software engineering research community.  


\section{Developer Survey}\label{sec:developer-survey}
We conduct the very first survey in the software engineering literature focusing on the practice of professional humor by developers.
We have distributed a questionnaire through the most active online communities of developers, and collected unique quantitative and qualitative gems about the role of humor in software development.

\subsection{Methodology}
The developer survey is based on an online questionnaire. 
First, the questionnaire introduces the context with examples of real-world software humor.
This is followed by questions asking the participant about their response to, and experiences with, humor in software.
The final section asks for demographic information. 
\revised{In order to minimize bias and reach a diverse audience,} we publish our questionnaire on multiple public channels for software developers: the \emph{AskHN} forum on Hacker News \cite{aniche2018modern}, a developer chat group on Discord, two developer groups on LinkedIn, and $6$ communities on Reddit, including \href{https://www.reddit.com/r/programminghumour/}{r/programminghumour}.
Additionally, we share the questionnaire with the developers we interviewed for the case studies.
The questionnaire was released online on August $21$, $2023$ and closed after five weeks.
Our supplemental package includes the questionnaire, all responses, and further details \footnote{\url{https://doi.org/10.5281/zenodo.8386612}}.

We received $125$ responses from participants whose experience with software development ranges from less than a year to up to $40$ years.
They work across various domains, such as software, finance, communication, and health.
Of the $125$ participants, $116$ (nearly $92.8$\%) write or review code at least once a week.

The first question is about the reaction of the respondent to humor within software development.
More than $93$\% of the respondents ($117$ of $125$) say that they would react positively, while $8$ would be neutral or negative.
This fully confirms our core hypothesis that humor is an important channel for developer engagement.

\subsection{Where Does Humor Belong in Code?}

\begin{figure}
\centering
\href{https://doi.org/10.5281/zenodo.8386612}{\includegraphics[width=\columnwidth]{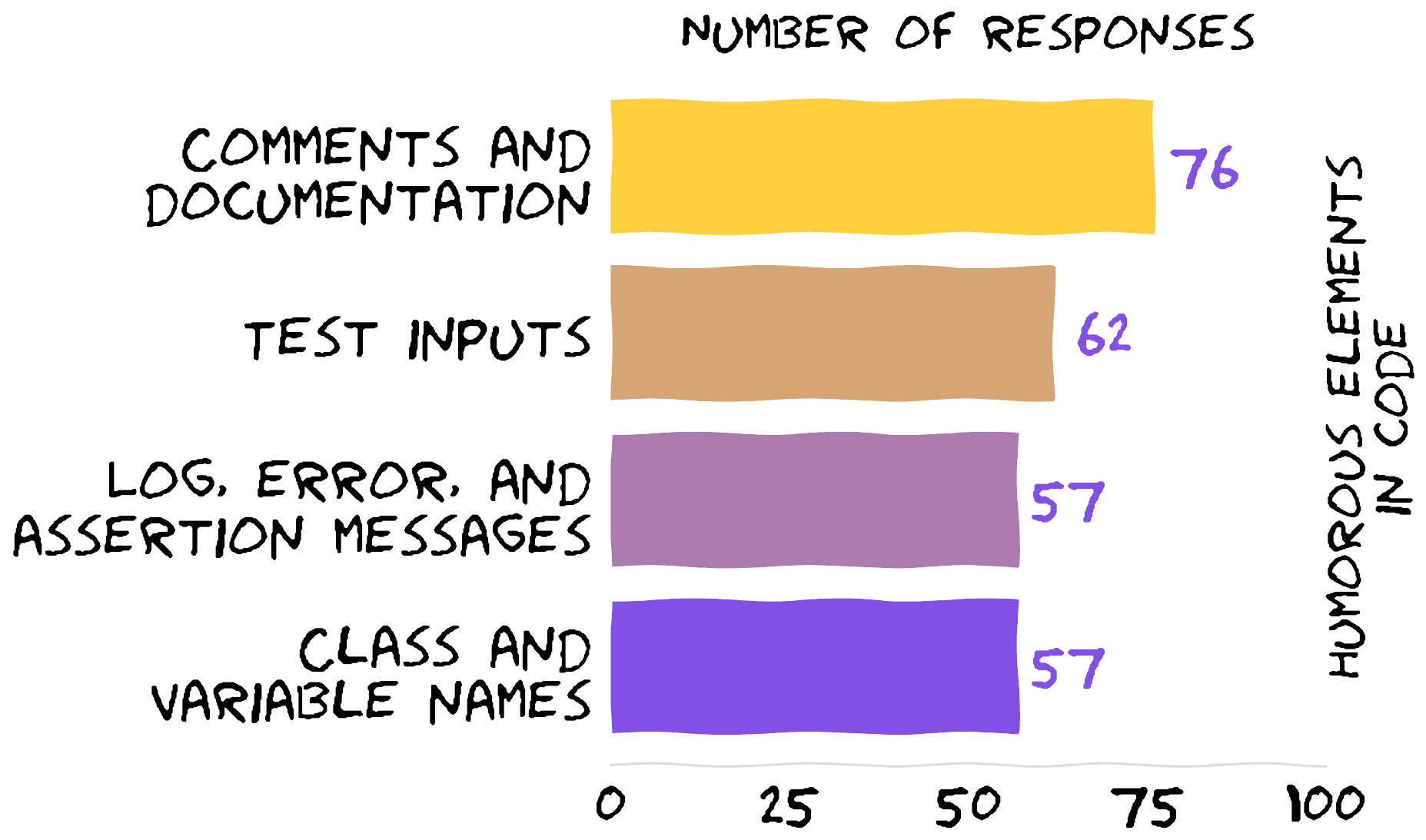}}
\caption{According to the survey respondents, the practice of humor in software is appropriate within  documentation, immediately followed by test inputs.}
\label{fig:survey}
\end{figure}

One important gap in the body of knowledge on software humor is about where it can be practiced.
Our survey contains a dedicated section about this.
The results are shown in 
\autoref{fig:survey}, which summarizes the top locations within software projects where developers engage with both code and humor.
Of the $100$ developers who reported having personally experienced software humor, $76$ say that \textbf{comments and documentation} are the best locations to unleash it.
Next comes a very special place in the heart of software humor practitioners: \textbf{test inputs}, with $62$ participants finding it appealing for humor.
This aligns with the \faker case study of \autoref{sec:faking-libraries}, showing that cultural references and fun quotes are appropriate test data.
Indeed, several participants mention \faker as one of the libraries that provides this opportunity.

\begingroup
\setlength{\tabcolsep}{12pt}
\renewcommand*{\arraystretch}{1.8}
\addstackgap[4pt]{
\begin{tabular}{!{\color{BurntOrange}\vrule width 2 pt}p{7.5cm}}
\emph{My humorous documentation is fully handcrafted.}\\
\end{tabular}
}
\endgroup

\begingroup
\setlength{\tabcolsep}{12pt}
\renewcommand*{\arraystretch}{1.8}
\addstackgap[4pt]{
\begin{tabular}{!{\color{BurntOrange}\vrule width 2 pt}p{7.5cm}}
\emph{I use test data to run wild with humor since it does really feel like the place for it.}\\
\end{tabular}
}
\endgroup

Per \autoref{fig:survey}, two other locations for the introduction of humor are within log messages, as well as the names of code elements such as variable, classes, or functions.
One developer shares their favorite example, the \texttt{phone\_home} function within the \texttt{et} module of the Erlang standard library \footnote{\url{https://www.erlang.org/doc/man/et.html}}.
Additionally, several developers mentioned other ways, outside of code elements, for sharing a joke with a colleague: within code review comments or inside Slack channels.

\begingroup
\setlength{\tabcolsep}{12pt}
\renewcommand*{\arraystretch}{1.8}
\addstackgap[4pt]{
\begin{tabular}{!{\color{BurntOrange}\vrule width 2 pt}p{7cm}}
\emph{Developers need to write \texttt{et.phone\_home} to use that  function, which is still funny to me.}\\
\end{tabular}
}
\endgroup

\begingroup
\setlength{\tabcolsep}{12pt}
\renewcommand*{\arraystretch}{1.8}
\addstackgap[4pt]{
\begin{tabular}{!{\color{BurntOrange}\vrule width 2 pt}p{7cm}}
\emph{I incorporate humor within library names. One library ensured only one process of a specific type was alive in a cluster. It was named Highlander, after the movie in which only one of the immortals was supposed to stay alive... ``There can be only one."}\\
\end{tabular}
}
\endgroup

The responses on our survey indicate that developers leverage humorous code to breathe life into their software projects.
They favor tests and documentation for placing humor.
From a social perspective, many developers admire humorous code authored by other developers.

Finally, the astute reader notices that we prepared the bar chart of \autoref{fig:survey} in \href{https://xkcd.com/}{xkcd} fashion using the dedicated \href{https://matplotlib.org/stable/api/_as_gen/matplotlib.pyplot.xkcd.html} {\texttt{matplotlib.pyplot. xkcd}} module.
Evidently, xkcd transcends the humorous comic strips; it is a source of inspiration for developers to create humorous, reusable software components such as this library.

\subsection{Humor, What is it is Good For?}
The key hypothesis of our work is that humor is fundamentally good in software development, to engage teams involved in hard engineering work. 
Our survey respondents highlight several compelling reasons for including humor within software development.

First, humor is oftentimes enjoyed for its own sake.
Developers solve technical challenges everyday, and humor can be be \textbf{a cheerful aid}.
Humor in code \emph{``helps keep it fun,"} shares one developer.

\begingroup
\setlength{\tabcolsep}{12pt}
\renewcommand*{\arraystretch}{1.8}
\addstackgap[4pt]{
\begin{tabular}{!{\color{BurntOrange}\vrule width 2 pt}p{7cm}}
\emph{Humor should be there. It really makes my day and makes me smile. I love it and think fondly of people writing that part of the code or comment.}\\
\end{tabular}
}
\endgroup

\begingroup
\setlength{\tabcolsep}{12pt}
\renewcommand*{\arraystretch}{1.8}
\addstackgap[4pt]{
\begin{tabular}{!{\color{BurntOrange}\vrule width 2 pt}p{7cm}}
\emph{I sometimes listen to an \href{https://open.spotify.com/album/3Dw1bhaOQxGiAD9IUt5grJ}{album} from a friend which has some fun track names, like ``Callback Hell" and ``Merge Conflict." It makes my day lighter.}\\
\end{tabular}
}
\endgroup

Second, humor promotes \textbf{a sense of community}.
For example, one participant recalls feeling connected to, and appreciative of, past and current colleagues who have authored humorous test inputs.
One participant also mentions that humor can be an ice-breaker, especially for remote collaborators.

\begingroup
\setlength{\tabcolsep}{12pt}
\renewcommand*{\arraystretch}{1.8}
\addstackgap[4pt]{
\begin{tabular}{!{\color{BurntOrange}\vrule width 2 pt}p{7cm}}
\emph{It makes a codebase feel more humanized, like it was created by a real person.}\\
\end{tabular}
}
\endgroup

\begingroup
\setlength{\tabcolsep}{12pt}
\renewcommand*{\arraystretch}{1.8}
\addstackgap[4pt]{
\begin{tabular}{!{\color{BurntOrange}\vrule width 2 pt}p{7cm}}
\emph{My cat named Bob showed up in a call and a colleague asked: ``oh, so that's the Bob Cat I've seen in the tests?" LOL.}\\
\end{tabular}
}
\endgroup

In addition to these positive effects, humor can have a practical impact on software development. 
For example, humor in test cases can facilitate \textbf{a common understanding} of the test intention. Funny stories that are invented and shared among colleagues create bonds that are beneficial when teams face adversity at the software factory.

\begingroup
\setlength{\tabcolsep}{12pt}
\renewcommand*{\arraystretch}{1.8}
\addstackgap[4pt]{
\begin{tabular}{!{\color{BurntOrange}\vrule width 2 pt}p{7cm}}
\emph{I incorporate something light (vs serious) to make things more fun and a bit easier to understand.}\\
\end{tabular}
}
\endgroup

\begingroup
\setlength{\tabcolsep}{12pt}
\renewcommand*{\arraystretch}{1.8}
\addstackgap[4pt]{
\begin{tabular}{!{\color{BurntOrange}\vrule width 2 pt}p{7cm}}
\emph{Inventing personas within integration tests seems to have rubbed off in my team and we now reference them when discussing the testing approach for new features. It helps lighten the mood and make some boring tasks a bit fun.}\\
\end{tabular}
}
\endgroup

Overall, our survey respondents confirm that humor is a good way of increasing developer engagement.
Humor makes the engineering process exciting, lightens the mood, promotes bonding, and may even contribute to more understandable code.

\subsection{Responsible Humor?}
With great humor, comes great responsibility. 
Developers balance the benefits of humor with other aspects of professional software development.
Some respondents of our survey avoid incorporating humor within software, because they think that it may be detrimental to its quality.

\begingroup
\setlength{\tabcolsep}{12pt}
\renewcommand*{\arraystretch}{1.8}
\addstackgap[4pt]{
\begin{tabular}{!{\color{BurntOrange}\vrule width 2 pt}p{7cm}}
\emph{I actively discourage my teams from getting cute because it often reduces readability and knowledge transfer.}\\
\end{tabular}
}
\endgroup

\begingroup
\setlength{\tabcolsep}{12pt}
\renewcommand*{\arraystretch}{1.8}
\addstackgap[4pt]{
\begin{tabular}{!{\color{BurntOrange}\vrule width 2 pt}p{7cm}}
\emph{Adding extra dependencies for humor would be a big no for me.}\\
\end{tabular}
}
\endgroup

Meanwhile, many developers take a more nuanced approach to humor within software, and recommend exercising caution when practising it.
For example, the location for the introduction of humor within code should be chosen wisely, considering its \textbf{impact on performance and quality}.

\begingroup
\setlength{\tabcolsep}{12pt}
\renewcommand*{\arraystretch}{1.8}
\addstackgap[4pt]{
\begin{tabular}{!{\color{BurntOrange}\vrule width 2 pt}p{7cm}}
\emph{Code functionality should not be affected by humor, but documentation or comments are a good place for that whenever appropriate.}\\
\end{tabular}
}
\endgroup

\begingroup
\setlength{\tabcolsep}{12pt}
\renewcommand*{\arraystretch}{1.8}
\addstackgap[4pt]{
\begin{tabular}{!{\color{BurntOrange}\vrule width 2 pt}p{7cm}}
\emph{I avoid writing humorous code that could be exposed to the end user, and ensure that my humorous tests are still straightforward and readable. No jokes just for the sake of jokes.}\\
\end{tabular}
}
\endgroup

For a responsible practice of humor, it is important to be mindful of the audience.
Humor should not hinder \textbf{respectful collaboration}.
Indeed, code review is important for risk management in humor, as highlighted by Stefanni (cf. \autoref{sec:faking-libraries}).
\emph{``I ensure that no contribution to \faker is offensive, which can be hard to do with generators like \texttt{south\_park}."}

\begingroup
\setlength{\tabcolsep}{12pt}
\renewcommand*{\arraystretch}{1.8}
\addstackgap[4pt]{
\begin{tabular}{!{\color{BurntOrange}\vrule width 2 pt}p{7cm}}
\emph{Humor in software projects, especially pop-culture references, should only be incorporated if you are cautious and diligent. It should not create a toxic or unwelcoming culture.}\\
\end{tabular}
}
\endgroup

To sum up, our survey respondents offer actionable advice, and many aspects discussed here could readily be adapted to fit in the ``Code of Conduct" or ``Contribution guidelines" of developer communities.
Clearly, humorous code should be respectful, in line with the global social responsibility theory \cite{christians2004social}.

\subsection{Summary}

Our three case studies and survey enable us to provide guidelines for developers and managers who want to introduce or encourage humor within their team:
\begin{itemize}
    \item Code comments are perfect for expressing humanity and humor, with guarantees that humor does not impact the final product. For example, the comments in the Apollo 11 source code authored by Margaret Hamilton and her crew did not prevent a successful moonshot, as we see from \autoref{lst:apollo}.

    \item Test inputs are appropriate for humorous creativity in code, and developers can rely on state of the art open source libraries to support this practice in different technical stacks.  
    
    \item The more challenging the development task, the more tech leads should resort to humor as a facilitator of communication and collaboration, see \autoref{lst:apollo}. 
\end{itemize}

\begin{mdframed}[nobreak=true,style=mpdframe, frametitle=Summary of the developer survey]
The developers surveyed across different continents and domains admit to expressing a bit of themselves through software humor.
They highlight that a responsible practice of software humor does not adversely impact code quality and the community.
The introduction of humor within software documentation and tests is an organic way to positively engage developers.
\end{mdframed}

\section{Related Work on Humor in Software}\label{sec:related-work}

We now discuss related forms of software humor, which has many different facets.

\emph{Software Easter Eggs}. They are one of the oldest forms of humor in software \cite{salvador2017history}.
This form of software humor consists of small pieces of code, hidden deep within commercial applications, meant to be fun and surprising.
Over the decades, Easter eggs have been creative interludes for developers and true enchantment for users.
Today, it is possible to find software Easter eggs in all sectors \cite{lakier2022more}, from video games \cite{whale18} to cars \cite{saldner2020design}, and web browsers \cite{baudry2022long}. 
The smooth acceptance and the longevity of the egg require operational software engineering excellence \cite{limoncelli2018operational}.
In particular, testing Easter eggs is notoriously hard \cite{xie2008using}.
Cacciotto and colleagues suggest that Easter eggs in graphical interfaces encourage software testers to explore rare behaviors of the interface \cite{9440184}.
Ruus evaluated how the Konami Easter egg can be used to detect cross-browser compatibility bugs \cite{ruushunting}.
Overall, this related literature shows that Easter eggs are positive intakes that favor developer engagement.

\emph{Humor and Software Development Education.}
Various studies have investigated the role that humor can play in education \cite{christman2018instructor}.
In particular, these studies look at humor to engage students with difficult topics \cite{chabeli2015humour}, or subjects perceived as boring \cite{jones2014humor}.
Jones documented the social, communication, and cognitive benefits of humor in education \cite{jones2014humor}.
In relation with software, Vogler and colleagues have demonstrated the strong interactions between learning and humor in online teaching \cite{vogler2019lolsquared}.
Takbiri and colleagues succeeded in introducing a gamification element to engage with K-6 students and increase their motivation \cite{takbiri2023gamified}.
As for software engineering education,  Prasetya and colleagues used various gamification techniques to teach formal methods \cite{PrasetyaLMTBEKM19}.
Jwo made software engineering education a joyful learning experience using comics to convey the software development lifecycle \cite{jwo2015teaching}.
Henry Gardner composed a series of humorous songs that reflect the different phases of software development, and has shown that these songs improve the appreciation of software engineering among his students \cite{gardnerhumorous}.
\_why\_the\_lucky\_stiff wrote a book about Ruby, heavily relying on humor and cartoons, to make programming more appealing to adolescents \cite{why}.

\emph{Pataphysics, Esolangs, etc.} The more we search for it, the more we find it. 
Raczinski explored pataphysics to develop a  search engine that returns humorous and provocative search results instead of purely relevant ones \cite{raczinski2013creative}.
Rick-rollette is a Chromium browser extension that Rickrolls its user with a 1\% probability per click \footnote{\url{https://github.com/DaviAMSilva/Rick-Rollette}}.
Esoteric languages, aka Esolangs  \cite{temkin2017language}, are programming languages with no direct utility except for humor  and curiosity \cite{esolangs}.
The authors and users of these languages have fun with the core design of programming languages \cite{Mateas08}.
In the past decade, emoji have become an essential element of conveying humor in software development, on GitHub \cite{wang2023react} and Stack Overflow \cite{venigalla2021stackemo}.
Clearly, humor is a piece in the grand puzzle of the pursuit of developer happiness, a key factor for code quality and software creativity \cite{graziotin2014happy}.


\section{Conclusion}\label{sec:conclusion}
This paper has explored the different ways in which humor is eating software.
We have drawn attention to three software projects that are appreciated by developers because of their humorous nature.
Libraries such as \faker eliminates boredom and invokes creativity within testing.
\lolcommits facilitates the celebration of each contribution made to a project by capturing developer photographs.
Useful technology can arise from software made purely for fun, as we have seen with the \volkswagen project.
When handled with care and responsibility, humor is meritorious and inclusive.
Real-world practitioners incorporate it within the software that powers all sectors of modern society. 
Humor promotes a sense of community within colleagues, and serves as an outlet for self expression.

 \begin{lstlisting}[style=ruby, showlines=true, numbers=none, label={lst:apollo}, belowskip=-2em, caption={Excerpt from the \href{https://github.com/chrislgarry/Apollo-11/blob/master/Luminary099/BURN\_BABY\_BURN--MASTER\_IGNITION\_ROUTINE.agc\#L45}{ignition routine} in the Apollo 11 code that sent humans \href{https://github.com/faker-ruby/faker/blob/2f06350da484de1fbc3f2e8ef3f1c755502a7a45/lib/locales/en/futurama.yml\#L204}{to the moon}. When Margaret Hamilton and her crew developed the Apollo Guidance Computer, they were also fueled by humor \cite{apollo11}. Software humor predates the invention of ``software engineering'' as a concept at the 1968 NATO conference \cite{naur1969software}.},float]

# BURN, BABY, BURN -- MASTER IGNITION ROUTINE
  BANK	36
  SETLOC	P40S
  BANK
  EBANK=	WHICH
  COUNT*	$$/P40

# THE MASTER IGNITION ROUTINE IS DESIGNED FOR USE BY THE
# FOLLOWING LEM PROGRAMS:  P12, P40, P42, P61, P63.
...
# HONI SOIT QUI MAL Y PENSE
...
\end{lstlisting}

Our study barely scratches the surface of developer humor, and there is yet much to unearth.
For example, research is needed to study the prevalence of memes in software engineering~\cite{iloh2021culture}. We have recently received a grant to investigate the dynamics of humor specialization per software stack, grounded on preliminary results on Rust  \cite{rustapril} and past research on humor niches in French elite education \cite{Verschueren15}.


\revised{\emph{The Software Humor Regimen. }
Based on our study, developers and researchers can introduce more humor in software, per the following responsible regimen.
Thou shalt embrace humorous libraries and naming conventions for serious test suites.
Thou shalt celebrate creative contributions to git repositories.
Thou shalt relish the serendipitous nature of open source development.
Thou shalt harness the mystique of humor to sustain healthy collaborative software development.
Thou shalt abide by the legacy of Margaret Hamilton and her crew who sent humorous code comments to the moon.
}

\section*{Acknowledgments}
We thank the interviewees, prolific humor and software practitioners: Thomas Watson, Fania Raczinski, Stefanni Brasil, Selena Small, Michael Milewski, Evan Travers, and Dawid Dylowicz.
We are deeply grateful to all the participants of our survey for sharing their experiences with us. 
\revised{This work has been supported by the Wallenberg Autonomous Systems and Software Program (WASP), funded by the Knut and Alice Wallenberg Foundation.}

\balance
\bibliographystyle{unsrteasteregg}
\bibliography{main}

\end{document}